\documentclass[12pt]{article}
\title{An Attempt to Design a Better Algorithm for the Uncapacitated Facility Location Problem}
\author{Haotian Jiang}
\usepackage{amsmath,amsfonts,latexsym,graphicx}
\usepackage{fullpage,color}
\usepackage{bm}
\usepackage{multicol}
\usepackage{url,hyperref}
\usepackage[linesnumbered,boxed,ruled,vlined]{algorithm2e}

\newtheorem{lemma}{Lemma}

\newtheorem{defn}[lemma]{Definition}

        {\medskip}

        {\hspace*{\fill}$\Box$\par}
        {\hspace*{\fill}$\Box$\par\vspace{4mm}}
        {\hspace*{\fill}$\Box$\par}

\newcommand{\E}{\textbf{E}}

\newcommand{\dd}{\mathrm{d}}

\begin{document}
\maketitle

\begin{abstract}
The uncapacitated facility location has always been an important problem due to its connection to operational research and infrastructure planning. Byrka obtained an algorithm that is parametrized by $\gamma$ and proved that it is optimal when $\gamma>1.6774$. He also proved that the algorithm achieved an approximation ratio of 1.50. A later work by Shi Li achieved an approximation factor of 1.488. In this research, we studied these algorithms and several related works. Although we didn't improve upon the algorithm of Shi Li, our work did provide some insight into the problem. We also reframed the problem as a vector game, which provided a framework to design balanced algorithms for this problem. 
\end{abstract}

\section{Introduction}
\qquad The uncapacitated facility location (UFL) problem has received lots of attention due to its connection to operational research and infrastructure planning.  In the UFL problem, we are given a set of potential facility locations $F$, each $i\in F$ with a facility cost $f_i$, a set of clients $C$ and a metric $d$ over $F\cup C$. We need to open a set of facilities $F'$, and assign each client to an opened facility, so that the sum of the total facility opening cost and the total assignment cost is minimized.

The UFL is NP-hard and the current best algorithm for this problem was designed and analyzed by Li \cite{ShiLi}, that reached an approximation ratio of 1.488. Another algorithm by Byrka, parametrized by $\gamma$ achieved an optimal bi-factor approximation ratio when $\gamma>1.6774$. All these algorithms are based on the algorithm of Chudak and Shmoys \cite{CS} (known as the Chudak-Shmoys algorithm) and a filtering technique of Lin and Vitter\cite{Lin}. The algorithms of Byrka and Li also utilized a (1.11,1.78) bifactor algorithm by Jain et al.\cite{Jain1}. 

On the negative side, Guha and Kuller \cite{Guha} showed that there is no 1.463-approximation unless \textbf{NP}$\subseteq$\textbf{DTIME}$(n^{O(\log\log n)})$. This condition was later strengthened by Sviridenko \cite{Svir} to be "\textbf{NP}=\textbf{P}". Jain et al. \cite{Jain1} generalized these work and proved that there is no $(\gamma_f,\gamma_c)$ bifactor approximation algorithm for $\gamma_c<1+2e^{-\gamma_f}$ unless  \textbf{NP}$\subseteq$\textbf{DTIME}$(n^{O(\log\log n)})$. We therefore refer to this result as the hardness curve of the UFL problem (notice that the hardness curve pass the point (1.463,1.463)). The algorithm by Byrka \cite{Byrka1}\cite{Byrka2} was optimal for $\gamma_f>1.6774$, but for $\gamma_f<1.6774$, there is no current algorithm that can reach this hardness curve.

The main difficulty in tackling the problem is to understand the regularity and irregularity of the instance. A instance is said to "regular" if all facilities to which a client is served in the optimal fractional solution to the LP, which is shown below, is of approximate the same distance to that client. While in an irregular case, the facilities to which a client is served in the optimal fractional solution varies in distance to that client. We have to notice that irregularity is not an absolute definition. The degree to which the instance is irregular may be different for different instances. We now consider an adversary that might produce instances to fight against our algorithms. It was shown that to fight against Byrka's algorithm for $\gamma>\gamma_0$ \ref{Byrka} or Li's algorithm\ref{Li}, the adversary would like to produce an completely regular case. In the case of Byrka's algorithm, that means that the case when $d_{ave}^C(j)=d_{ave}^D(j)$ will make the bounds tight. In the case of Li's algorithm, that means that the best strategy by an adversary would be render the characteristic function a threshold function. On the other hand, however, the regular case is also the easiest case to deal with, as is indicated by section 3.1 in Byrka's paper \cite{Byrka2}. Our failing to come up with an algorithm that does better than that of Li's indicates the existence of a worst case to Li's algorithm that is a very irregular case.

This research serves as an attempt to improve over the 1.488-approximation algorithm of Li. He allowed the algorithm to make a random choice of $\gamma$, which was the parameter in Byrka's algorithm, and in combination with the $(1.11,1.78)$ JMS algorithm, obtained the 1.488-approximation algorithm. In his paper, he developed a "zero-sum game" whose value equals the approximation ratio of the algorithm and found the best strategy for the adversary that made it easy to analyze the algorithm. This best strategy turned out to be a very "regular" instance which was rather easy to handle if the parameter $\gamma$ was chosen optimally instead of randomly. We also observed that the so-called "zero-sum game" was in fact not a "zero-sum game". Instead, it was a vector game at its core and thus the Minimax Principle didn't apply. Based on these observations, we attempted to improve the algorithm by making an optimal choice of $\gamma$ instead of a random choice. Our analysis of the algorithm used Blackwell's Approachability Theorem. Unfortunately, our computer simulation showed that our algorithm did not  actually improve Li's algorithm. However, this failure might bring more insights into either the UFL problem or the theory of vector games.
 
\section{Byrka's Algorithm and Analysis}
\label{Byrka}
\quad In this section, we review the algorithm by Byrka\cite{Byrka2}, yet our notation follows that of Li's\cite{ShiLi}. The algorithm starts by solving the following linear programming for the UFL problem:
\begin{eqnarray*}
\min &&\quad \sum_{i\in F, j\in C} d(i,j)x_{i,j}+\sum_{i\in F}f_iy_i\quad s.t.\\
&&\quad \sum_{i\in F}=1\qquad \forall j\in C\\
&&\quad x_{i,j}\leq y_i\qquad \forall i\in F,j\in C\\
&&\quad x_{i,j},y_i\geq 0 \qquad \forall i\in F,j\in C
\end{eqnarray*}
Throughout this article, we will assume that we have solved the above LP and obtain an optimal solution $x^*$ and $y^*$. For each client $j$, the set of facilities to which it is served more than 0 is denoted as $F_j$.

The algorithm starts by a filtering process with parameter $\gamma$: we create a "filtered" solution $\overline{x}$, $\overline{y}$ as follows: $\overline{y}=\gamma y^*$ . We may split those facilities with $\overline{y}>1$ to two facilities. Then since the $\overline{y}$ are fixed, we may reassign the $\overline{x}$ facilities in a greedy manner: for each client $j$, sort the facilities by their distances to $j$ and then for each facility $i$ in order, assign $\overline{x}_{ij}=\overline{y_i}$ until the sum of all assignments of client $j$ reaches 1. W.l.o.g., we may assume that the solution obtained in this way is a complete solution, that is $\overline{x}_{ij}$ is either $\overline{y_i}$ or 0. This can be done via a facility splitting argument. A more detailed one can be found in \cite{}.

For each client $j\in C$, after the filtering process, we may distinguish between two kinds of facilities with respect to that client: close facilities and distant facilities. We say a facility $i$ is a close facility to client $j$ if $\overline{x}_{ij}>0$. We say a facility $j$ is a distant facility to client $j$ if $\overline{x}_{ij}=0$ but $x^*_{ij}>0$. We denote the set of close and distant facilities to client $j$ as $F_j^C$ and $F_j^D$ respectively.

Now we define a distance from a client $j$ to a set of facilities $F'\subseteq F$ $d(j, F')$ as the average distance to the facilities $F'$ with respect to the solution $\overline{y}$. We define $d_{ave}^C(j)=d(j,F_j^C)$, $d_{ave}^D(j)=d(j,F_j^D)$ and $d_{ave}(j)=d(j,F_j)$. We also define $d_{max}^C(j)$ as the maximum distance from $j$ to a facility in $F_j^C$. 

Byrka's algorith follows the idea developed Chudak and Shmoys, which first select some facilities as the cluster centers, according to some criteria quantity. However, the criteria quantity chosen here is $K(j)=d_{max}^C(j)+d_{ave}^C(j)$. The process of choosing clustering centers is as follows. We consider each client in the order of increasing criteria quantity $K$. Suppose we are now at client $j$, if it is not assigned to any clustering center, we make it a clustering center, and for all those clients $j'$ that satisfies $F^C_{j'}\cap F^C_j\neq \emptyset$, we assign client $j'$ to the cluster center $j$. After this process, the clients are divided into several clusters. Call the set of all cluster centers $C'$.

We now round the solution $(\overline{x},\overline{y})$ randomly. For each cluster center, we open exactly one of its close facilitis randomly with probabilities $\overline{y}_i$. For each facility that is not a close facility of any cluster centers, open it independently with probability $\overline{y}_i$. Then connect each client $j$ to its closest facility. Let $C_j$ be the connection cost of client $j$.

The expected cost of the algorithm is just $\gamma$. The key to bounding the expected total connection cost is to bound $C_j$ under the condition that there is no open facility in $F_j$. Byrka was able to show that $d(j,F_{j'}^C\backslash F_j)\leq d_{ave}^D(j)+d_{max}^C(j)+d_{ave}^C(j)$ and thus bound the $\E[C_j]\leq (1-e^{-1}+e^{-\gamma})d_{ave}^C(j)+(e^{-1}+e^{-\gamma})d_{ave}^D(j)$. But the connection cost of the optimal solution is $d_{ave}(j)=\frac{1}{\gamma}d_{ave}^C(j)+\frac{\gamma-1}{\gamma}d_{ave}^D(j)$. By computing the maximum ratio between these two costs, we have an optimal bifactor algorithm for $\gamma>\gamma_0\approx 1.67736$ that achieves a bifactor approximation ratio of $(\gamma,1+2e^{-\gamma})$. Also by taking $\gamma=\gamma_0$ and in combination with the $(1.11,1.7764)$-bifactor approximation by Jain, Mahadian and Saberi, we can get a 1.50-approximation algorithm for the UFL problem.

In his paper \cite{Byrka2}, he also pointed out that by making $\gamma$ random instead of a fixed value might improve the approximation ratio and that's the basis of Li's algorithm and analysis.
\section{Li's Algorithm and Analysis}
\label{Li}
Li's algorithm is based on Byrka's algorithm by making $\gamma$ random. Yet the analysis is very involved.

Li's analysis utilized an important concept called characteristic function (see Definition 15 in Li's paper \cite{ShiLi}.
\begin{defn}[Definition 15 in Li's paper] Given a UFL instance and its optimal fractional solution $(x,y)$, the characteristic function $h_j:[0,1]\rightarrow \mathbb{R}$ of some client $j\in C$ is defined as follows. Let $i_1,i_2,\cdots,i_m$ be the facilities in $F_j$, in the non-decreasing order of distances to $j$. Then $h_j(p)=d(i_t,j)$, where $t$ is the minimum number such that $\sum_{s=1}^ty_{i_s}\geq p$. The characteristic function of the instance is defined as $h=\sum_{j\in C}h_j$.
\end{defn}
It is not hard to see that the characteristic function is a non-decreasing piece-wise constant function. Li also came up with a better bound on the connection cost, which is Lemma 12 in his paper \cite{ShiLi}.
\begin{lemma}[Lemma 12 in Li's paper]
For some facility $j\notin C'$, let $j'$ be the cluster center of $j$. We have,
$$
d(j,F_{j'}^C\backslash F_j)\leq (2-\gamma)d_{max}^C(j)+(\gamma-1)d_{ave}^D(j)+d_{max}^C(j')+d_{ave}^C(j').
$$
\end{lemma} 
With this bound, he was able to show the following bound on the expected connection cost.
\begin{lemma}[Lemma 20 in Li's paper]
The expected cost of the integral solution is 
$$
\E[C]\leq \int_0^1h(p)e^{-\gamma p}\gamma \dd p+e^{-\gamma}(\gamma\int_0^1h(p)\dd p+(3-\gamma)h(\frac{1}{\gamma})).
$$
\end{lemma}
To come up with an explicit distribution for $\gamma$, Li introduced a zero-sum game. We can scale the instance so that $\int_0^1h(p)\dd p=1$. Let 
$$\alpha(\gamma,h)=\int_0^1h(p)e^{-\gamma p}\gamma \dd p+e^{-\gamma}(\gamma+(3-\gamma)h(\frac{1}{\gamma}))$$. 
The game is between an algorithm designer $A$ and an adversary $B$. The strategy of $A$ is a pair $(\mu,\theta)$, where $\theta$ is the probability of playing the JMS algorithm and $\mu$ is $1-\theta$ times a probability density function for $\gamma$. The strategy for $B$ is a non-decreasing piece-wise constant function $h:[0,1]\rightarrow \textbf{R}*$ such that $\int_0^1h(p)\dd p=1$. The value of the game, when the strategy of $A$ is $(\mu,\theta)$ and the strategy of $B$ is $h$, is defined as:
$$
\nu(\mu,\theta,h)=\max\{\int_1^\infty \gamma\mu(\gamma)\dd \gamma+1.11\theta,\int_1^\infty \alpha(\gamma,h)\mu(\gamma)\dd \gamma+1.78\theta\}
$$
A threshold function $h_q:[0,1]\rightarrow \mathbb{R},0\leq q < 1$ is defined as:
$$
h_q(p)=
\begin{cases}
0 &\quad p\leq q\\
\frac{1}{1-q} & \quad p>q
\end{cases}
$$

And Li was able to show that there is a best strategy for $B$ that is a threshold function. From this fact he was able to find the optimal strategy for $A$: with probability $\theta\approx 0.2$, run the JMS algorithm; with probability about 0.5, run Byrka's algorithm witth parameter $\gamma_1\approx 1.5$; with the remaining probability, run Byrka's algorithm with parameter $\gamma$ selected uniformly between $\gamma_1$ and $\gamma_2\approx 2$. A very refined analysis by Li proved that this distribution was able to achieve an approximation ratio of 1.4879.
\section{An Attempt to Improve Li's Algorithm}

 \quad In this section, we present our attempt to improve over Li's algorithm. Although computer simulation illustrates that this approach might not actually work, it does provide some insights. It also provides a framework to design better approximation algorithms.

Li's analysis showed that there is a best strategy for $B$ which is a threshold function, which corresponds to a very regular case. But if the instance is regular, we might not need to randomly pick $\gamma$. It might be better if we can examine the instance carefully and find the best $\gamma$ for the instance. This idea makes use of the fact that there is an order in the game: the adversary first pick an instance and then the algorithm designer runs his algorithm. That means the algorithm designer is able to first examine the instance and then make his dicision in choosing the best parameter $\gamma$, which might help to improve the approximation ratio.

Our method of choosing the best parameter $\gamma$ is as follows: first discretize the region $[1,M]$, where $M$ is a constant large enough, to many possible choices for the parameter $\gamma$. Then for each choice of $\gamma$, find the correspondingly best $\theta$ so that the balanced approximation ratio is the smallest. Then choose the $\gamma$ that corresponds to the smallest approximation ratio, and the corresponding $\theta$ that achieves it. 

It is obvious that this approach achieves a performance at least as good as Li's algorithm, since choosing the parameter randomly cannot do better than picking it optimally. Another very important reason why this approach might does better is due to the fact that the game defined by Li is not strictly a zero-sum game. According to Von Neumann's Minimax Principle, it will not be helpful if it is a zero-sum game, since a best strategy by $B$ will not be influenced by the order of the players. But the game is not a zero-sum game, since the value of the game is a maximum between two quatities. We shall now restate this game as a vector game. 

To make it easier to state and more importantly, easy to simulate with a computer, we shall apply discretization to the game. Notice the game is in fact continuous and its definition is almost the same with what we are going to present here. We shall first discretize the region in which $\gamma$ might take value to very tiny pieces and call these discretized value $\gamma_1,\gamma_2,\cdots$. These indicates the pure strategy of $A$ if he is going to run Byrka's algorithm. Another pure strategy of $A$ is the JMS algorithm. So the strategy set of $A$ can be written as $\{JMS,\gamma_1,\gamma_2,\cdots\}$, and $A$ is allowed to play a mixed strategy. The pure strategy of $A$ is denoted by $(\theta,\gamma)$, where $\theta$ is either 1 (indicating the JMS algorithm), or 0 (where in this case, $A$ runs Byrka's algorithm with parameter $\gamma$). Now we define the strategy of $B$. Since every characteristic function can be written as a linear combination of a finite number of threshold function, we might think of the strategy of $B$ as a mixed strategy over all threshold functions. We can therefore discretize the region $[0,1]$ to very tiny pieces $p_1,p_2\cdots$, and the strategy set of $B$ is thus $\{p_1,p_2,\cdots\}$. The value of the game is in fact a vector $\alpha$, which is defined as $\alpha(\theta, \gamma, p)=(C_f,C_c)(\theta,\gamma,p)=((1-\theta)\gamma+1.11\theta,(1-\theta)\alpha(\gamma,h_p)+1.78\theta)$. We should stress here that this value is defined when both players play pure strategy, thus $\theta\in \{0,1\}$. If players play mixed strategies, then we shall define the value of the game in an intuitive way, that is the expected vector value. The approximation ratio of this algorithm is determined by the region which is approachable by $A$. That is, we seek to find the smallest $\beta$ so that the 2-D region defined by the half-plains $x\leq \beta,y\leq \beta$ is approachable. In this case, $\beta$ will be the approximation ratio of this algorithm.

To find such a $\beta$, we may first guess how small can it be and we suppose our guess is $\beta_0$. Our task at this moment is to show that the region defined by $\beta_0$ is approachable. According to Blackwell's Approachability Theorem, it suffices to prove the approachability of all the supporting hyperplanes, which are just all the lines that pass through the point $(\beta_0,\beta_0)$ with non-positive slope in this case. We may assume that each of the line is defined by the equation with parameter $0\leq \phi\leq 1$: $\phi x+(1-\phi)y=\beta_0$, then if we only consider the game with respect to this line, we can get a zero-sum game with value $\phi C_f+(1-\phi)C_c-\beta_0$. We call this quantity the value of the game with respect to the line parametrized by $\phi$. When $\phi$ is determined, the remaining zero-sum game can be solved by any standard method, and in this research, we use linear programming to handle it. We seek to find a mixed strategy for $B$ such that for any pure strategy of $A$, the value of the game with respect to the line parametrized by $\phi$ is larger than 0. If we can find such a strategy for $B$, then the line parametrized by $\phi$ is not approachable. In the language of LP, the line parametrized by $\phi$ is not approachable if and only if the corresponding LP stated above has at least a solution that corresponds to a mixed strategy for $B$. We may therefore write out the LP explicitly and test whether or not a solution can be found.

If our guess for $\beta_0$ is indeed correct (notice it might not be the smallest of such correct guesses), Blackwell's Approachability Theorem tells us that for any $\phi\geq 0$, the line parametrized by $\phi$ is approachable. On the other hand, if we can check through all possible $0\leq\phi\leq 1$ and show that every one of them is approachable, we might conclude that the region defined by $\beta_0$ is approachable. To make the problem tractable to our computer, we should discretize for $\phi$ as well. Then we enumerate each possible value for $\phi$ and see if it is approachable.

In our simulation, we discretized each of $\gamma$,$p$ and $\phi$ to 1000 possible values and tested for approachability with different $\beta$. Unfortunately, the smallest such $\beta$ found was $1.4880$, which showed that our attempt to improve over Li's algorithm didn't work. However, the result found by this simulation is interesting, the startegy for $B$ turned out to be very complicated: it had a value of around 0.6 for $p=0$, and very tiny values ($\leq 0.02$) for all other values of $p$. But we have to be careful here in understanding this result because the coefficients of each $h_p$ is determined by the following equation: 
$$
h=\sum_{i=0}^{m-1}(c_{i+1}-c_i)(1-p_i)h_{p_i}
$$
Therefore, what the result really says is that there is an average distance of 0.6 where there are no facilities within it, but outside this region, facilities seem to be distributed very uniformly (in respect of distance) around the client.
\section{Implication of Our Results}
Although our attempt failed to improve over Li's algorithm, it had interesting implications to both the UFL problem and vector games.
\subsection{Implication to the UFL problem}
The result of our linear programming showed the existence of a highly irregular instance which could defeat the filtering process and our attempt to take the best $\gamma$ value. In other words, the filtering process alone cannot capture every aspect of the irregularity of the instance. If we want to design a better algorithm, we might need a better method to capture the irregularity of the problem.
\subsection{Implication to vector games}
As a byproduct, another interesting aspect of the implication of our result lies in the theory of vector games. At present, no corresponding result to the Minimax theorem of zero-sum games is known to vector games. However, our failure tends to show that the order of playing for vector games might not be important, under some condition. Admittedly, the vector game in our research is special in that (1) the first coordinate has nothing to do with the strategy of $B$ and (2) we only care about the maximum of the two coordinates. At this point, we don't know what aspects of this game leads to the fact that the order of playing is not important. Neither do we know whether a general conclusion exists for vector games. 
\section{Acknowledgements}
I thank my instructor Li for helpful discussions. 	

\bibliographystyle{plain}

\begin{thebibliography}{99}
\bibitem{Byrka1} Jaroslaw Byrka. An optimal bifactor approximation algorithm for the metric uncapacitated facility location problem. In \emph{APPROX ’07/RANDOM
’07: Proceedings of the 10th International Workshop on Approximation
and the 11th International Workshop on Randomization, and Combinatorial Optimization. Algorithms and Techniques}, pages 29-43, Berlin, Heidelberg, 2007. Springer-Verlag.
\bibitem{Byrka2} Jaroslaw Byrka, MohammadReza Ghodsi, and Aravind Srinivasan. LP-
rounding algorithms for facility-location problems. \emph{arxiv1007.3611}, 2010.
\bibitem{Guha} Sudipto Guha and Samir Khuller. Greedy strikes back: Improved facility
location algorithms. In \emph{J. Algorithms}, pages 649-657, 1998.
\bibitem{Jain1} Kamal Jain, Mohammad Mahdian, and Amin Saberi. A new greedy ap-
proach for facility location problems. In \emph{Proceedings of the thiry-fourth annual ACM symposium on Theory of computing}, STOC '02, pages 731-740, New York, NY, USA, 2002. ACM.
\bibitem{Jain2} Kamal Jain, Mohammad Mahdian, Evangelos Markakis, Amin Saberi, and
Vijay V. Vazirani. Greedy facility location algorithms analyzed using dual fitting with factor-revealing LP. \emph{J. ACM}, 50:795-824, November 2003.
\bibitem{ShiLi} Shi Li. A 1.488 approximation algorithm for the uncapacitated facility location problem. \emph{International Colloquium on Automata, Languages, and Programming}. Springer Berlin Heidelberg, 2011.
\bibitem{Lin} Jyh-Han Lin and Jeffrey Scott Vitter. Approximation algorithms for geo-
metric median problems. \emph{Inf. Process. Lett.}, 44:245-249, December 1992.
\bibitem{CS} Fabian A. Chudak and David B. Shmoys. Improved approximation algorithms for the uncapacitated facility location problem. \emph{SIAM J. Comput.},
33(1):1-25, 2004.
\bibitem{Svir} Maxim Sviridenko. An improved approximation algorithm for the metric
uncapacitated facility location problem, 2002.
\end{thebibliography}

\end{document}